\documentclass[12pt,a4paper]{article}
\usepackage[utf8]{inputenc}
\usepackage{fullpage}
\usepackage{authblk}
\usepackage[T1]{fontenc}
\usepackage{setspace}
\usepackage{etaremune}
\usepackage{amsmath}
\usepackage{mathpazo} 
\usepackage{graphicx}
\usepackage{hyperref}
\usepackage[english]{babel}
\usepackage{natbib}

\usepackage{fancyhdr}
\title{\textsc{Solar wind-driven day-to-day effects on the Martian thermosphere/exosphere composition}}
\author{\small{Kamsali Nagaraja$^{1}$, Praveen Kumar Basuvaraj$^{1}$, S.\,C. Chakravarty$^{1}$, K. Praveen Kumar$^{2}$}}
\affil{\textit{$^{1}$Department of Physics, Bangalore University, Bengaluru 560056 India}}
\affil{\textit{$^{2}$Space Science Programme Office, ISRO Headquarters, Bengaluru 560094 India}}

\date{}

\begin{document}

\maketitle
\vspace{-1.5cm}
\begin{center}
	Correspondance: Kamsali Nagaraja \href{mailto:kamsalinagaraj@bub.ernet.in}{(kamsalinagaraj@bub.ernet.in)} 
\end{center}

\hrulefill 

\begin{center}
	\large \underline{Abstract}
\end{center}

\textbf{Since the first \emph{in-situ} measurements of the altitude profile of upper atmospheric density and composition carried out by the Viking lander missions during 1976, similar data were continuously gathered by MAVEN and MOM spacecraft orbiting Mars since September 2014 with a mass spectrometer and other related payloads. Using near-simultaneous observations by the two orbiters, we show that both data sets confirm significant day-to-day variations of Argon ($Ar$) density profiles in the Martian thermosphere/exosphere during 1-15 June 2018 when the solar EUV radiation did not show any appreciable change. We extend this study to include the parent atmospheric constituents ($CO_{2}$, $Ar$, $He$, $N_{2}$) and the photochemical products ($O$, $CO$) to examine the effect of solar wind plasma ($e/H^{+}$) velocities and fluxes during the above time interval. Density profiles of these constituents show significant effects due to the additional electron impact dissociation and ionisation during the first week of June 2018, which subside in the next week returning to normal conditions. These first-time results are interpreted based on a number of relevant neutral and ion chemical reactions. This result provides a vital input to future modelling efforts of Martian thermosphere/exosphere composition studies and the solar EUV related variations due to the Schwabe cycle.}


\section*{Introduction}

Considerable progress has been made to conduct \emph{in-situ} observations of Martian surface and atmospheric parameters using orbiters, landers and rovers. Near-surface meteorological data has been analysed and consolidated in terms of diurnal, seasonal and inter-annual variations for more than 20 Martian years \citep{Martinez2017NearSurfaceMarsAtmos_VikingToCuriosity}. However, till recently, the measurements of upper atmospheric composition and density of Mars have been limited to the two sets of observations taken by the Viking landers while traversing down through its thin atmosphere \citep{Nier1976StrucOfMNUA_NMS_Viking1-2,NierHanson1976CompStruPrelimResult_NMS_Viking1,Owen1977CompOfMarsAtmosSurface} and hence it has not been possible to study its characteristic variations. \par

In September 2014, the Mars Atmosphere and Volatile Evolution (MAVEN) and the Mars Orbiter Mission (MOM) spacecraft entered into Martian orbit and placed in elliptical orbits around Mars with one of the main objectives to gather continuous data on spatial and temporal profiles of various upper atmospheric neutral/ion density and composition parameters \citep{Jakosky2015MAVEN,arunan2015MOM}. \par

Mars has a well-mixed region of the homosphere with the homopause at $\approx$120\,km altitude; the thermosphere is defined as the region above where the gases diffusively separate with individual gas species following their scale heights. This region finally merges into the exosphere, where the lighter gases may get energised to attain escape velocities. Usually, this loss process starts from around 220\,km, called the exobase, where the scale heights and mean free paths are comparable. The dynamics of this region is driven by the energy and momentum fluxes from planetary and tidal waves propagating through the lower atmosphere and by solar extreme ultraviolet (EUV) radiation, flare induced energetic particles and variable solar wind plasma fluxes \citep{Valeille2009MarsThermoIono3DModel_HotOxygenCorona,Medvedev2015CoolingMarsThermos_CompareGCM}. For Mars, the energetic charged particle interaction with atmospheric constituents, especially in the thermosphere-exosphere region, can change the electromagnetic radiation dominated dissociation and ion chemical reactions considerably due to the lack of any Earth-like magnetosphere.\par

The results obtained from both MAVEN using NGIMS (Neutral Gas and Ion Mass Spectrometer) payload and MOM using MENCA (Mars Exospheric Neutral Composition Analyser) payloads have so far provided new information about the spatial variation of the upper atmospheric gas constituents and ion species delineating their vertical and horizontal distributions and also the effect of the solar zenith angles with day-night, latitude and solar longitude variations \citep{Mahaffy2015NGIMS-MAVEN,Nagaraja2020ExosphereOfMars_MENCA-MOM}.\par

The primary purpose of this paper is to search and examine the effect of variation of solar wind plasma energetics on the short period day-to-day vertical distribution and density changes of different thermosphere-exosphere constituents based primarily on NGIMS/MAVEN data with near-simultaneous observations using MENCA/MOM for comparison.

\section*{Theory, Data and Method of Analysis}

Above the homopause, the individual atmospheric neutral species follow the altitude-density profile according to their molecular masses; hence heavier ones with shorter scale heights ($\boldsymbol{H^{*}}$) decrease rapidly  with altitude as given by the following relation derived from the condition of the hydrostatic equilibrium of the atmospheric pressure/density:

\begin{equation}
p_s(z)=p{_s}(z_0)\exp(-\frac{z-z{_0}}{H{^*}}) 	\label{eq001}
\end{equation}

where $\boldsymbol{p_{s}}$ is the pressure of species $\boldsymbol{s}$, $\boldsymbol{z}$ and $\boldsymbol{z_{0}}$ are heights and reference height respectively and, $\boldsymbol{H^{*}}$ is the scale height which is given by:

\begin{equation}
H{^*}=\frac{kT}{m{_s}g}  \label{eq002}
\end{equation}

where $\boldsymbol{k}$ is Boltzmann constant, $\boldsymbol{T}$ is the temperature, $\boldsymbol{m_{s}}$  the molecular mass of species $\boldsymbol{s}$ and $\boldsymbol{g}$ is the acceleration due to gravity.
\par

Like on Earth, the upper atmosphere of Mars gets ionised mainly by solar EUV radiation, and the ionosphere gets formed with the peak electron density around 120 km, varying with solar activity. As the electron-ion Coulomb collisions increase with height and the ion/electron-neutral collisions decrease, the electron and ion temperatures increase. The supra-thermal electrons produced by the EUV ionisation of neutral species above the ionisation peak height cause the enhanced plasma temperatures in this region above the homopause.\par

For Mars, another critical aspect is the direct interaction of solar energetic particle radiation or the non-thermal (as compared to electrons/ions in the ionosphere, which are thermalised through collisions) electrons and protons in the continuous flow of solar wind with the thermalised neutral and charged atmospheric species resulting in additional molecular dissociation, ionisation, recombination and charge exchange processes. These phenomena are critical to consider as they modify the equilibrium condition of the vertical distribution of neutral and charged species in the thermosphere-exosphere region and add to the escape of gases.\par

The electron impact dissociation and ionisation take place as in the following example:

\begin{eqnarray}
CO{_2} + e{^*} \rightarrow CO + O + e{^*} \label{eq003}\\
O{_2} + e{^*} \rightarrow O + O + e{^*} \label{eq004}\\
CO{_2} + e{^*} \rightarrow CO{_2^+} + e{^*} + e \label{eq005}\\
O{_2} + e{^*} \rightarrow O{_2^+} + e{^*} + e \label{eq006}\\
O + e{^*} \rightarrow O{^+} + e{^*} + e \label{eq007}
\end{eqnarray}

where $\boldsymbol{e{^*}}$ denotes energetic solar wind electron and $\boldsymbol{e}$ the thermalised electron as obtained from electron impact ionisation.\par

The electron impact ionisation frequency may be computed from the following equation: 

\begin{equation}
I{_e}{_i} = \int\sigma{_e}{_i}{(E)}\phi{_e}{(E)}dE \label{eq008}
\end{equation}

where $\boldsymbol{\sigma_{ei}}$ is the electron impact cross-section and $\boldsymbol{\phi{_e}}$ the solar wind electron flux; both being functions of energy $\boldsymbol{E}$. Similarly, the electron impact dissociation product densities can be estimated using appropriate reaction crossections. \par

Solar wind energetic protons enable charge exchange reactions by colliding with the exospheric constituents, which result in important changes in the exospheric distribution of neutral and ionic species, examples of which are given below:

\begin{eqnarray}
p{_s}{_w}{^+} + H \rightarrow H{^+} + H (fast) \label{eq009} \\
p{_s}{_w}{^+} + O \rightarrow O{^+} + H (fast) \label{eq010}
\end{eqnarray}

where $\boldsymbol{p{_s}{_w}{^+}}$ is the solar wind proton. \par

Knowing the charge exchange cross-section and the solar wind ion density and velocity, the production of new species can be estimated.\par

As mentioned earlier, here we search and explore the Martian thermosphere/exosphere composition data along with solar EUV and SW parameters obtained near-simultaneously from multiple payloads of MAVEN and MOM to study any effect of changing solar EM and charged particle radiation on the atmospheric density and composition mainly in thermosphere/exosphere region. The period June 1-15, 2018, has been selected for this study because with this MAVEN data, a few days of observation with the MENCA instrument is also available for any comparison. The selection of these days is also based on avoiding the second half of June 2018, which was affected by the Planet Encircling Dust Event (PEDE). The effect of the global dust storm on thermospheric densities has been studied using the available MENCA and NGIMS data for June 2018 and demonstrated the asymmetry between the daytime and nighttime observations of both the spacecraft \citep{Rao2020EnhanceDensity_MAVEN_MOM}. \citet{Bhardwaj2016EveningExosphereOfMars_MENCA-MOM} has given the instrumentation, calibration, sensitivity, and measurement limitations of the MENCA/MOM instrument. \par

The NGIMS instrument of the MAVEN spacecraft has been utilised to determine the structure and composition of the upper atmosphere's neutral and ionic species, including the isotopes and supra-thermal ions in a range of 2 to 150\,amu \citep{Mahaffy2015StrucCompMNUA_NGIMS-MAVEN}. NGIMS science-mode observation covers below 500\,km to periareion (inbound) and periareion to 500\,km (outbound) during each orbit lasting for $\approx$600\,s for each leg, with a spatial resolution of $\approx$1\,km. The level-2 data-sets of NGIMS-MAVEN have been retrieved from MAVEN Science Data Center to analyse and display results. \par

The archived MENCA data consists of total pressure and partial pressure values in Torr with a variable time resolution of $\approx$12 to 30\,s. This data is used for scientific studies after incorporating calibration and normalisation factors and time tagging with ancillary information like latitude, longitude, altitude, and solar zenith angle derived using  SPICE kernel files provided through the payload team of ISRO. The data sets are identified and arranged concerning different orbit numbers of MOM. \par

The Extreme UltraViolet Monitor (EUVM) instrument on MAVEN measures the solar irradiance at Mars using three photometers sensitive to the wavelengths  0.1-7\,nm, 17-22\,nm and 121.6\,nm. Apart from heating the Martian thermosphere/exosphere, the solar EUV radiation spectrum is responsible for the production and losses of chemical species through photodissociation, photoionisation, and suprathermal electrons \citep{Eparvier2015EUVM-MAVEN}. The EUV fluxes in the wavelength band 17-22\,nm have been analysed for the study period. \par

The Solar Energetic Particle (SEP) instrument on MAVEN consists of two dual, double-ended solid-state telescopes with four look directions per species, optimised for parallel and perpendicular Parker Spiral viewing of energetic ions (25\,keV to 12\,MeV) and electrons (25\,keV to 1\,MeV) with 1\,s time resolution. SEP can measure energy fluxes that range from 10 to 10$^{6}$ eV/(cm$^{2}$ s sr eV) to characterise solar energetic particles in an energy range that affects the upper atmosphere and ionosphere of Mars through sputtering, heating, dissociation, excitation, and ionisation. SEP has a mechanical pinhole attenuator that protects against overheating when the Sun is in its field of view \citep{Larson2015SEP-MAVEN}. SEP measured solar wind electron and $H$-ion velocities and densities during the same period cover the canonical Parker spiral directions around which solar energetic particle distributions are typically centred. The data of the hydrogen-ion velocities and fluxes have been analysed for the same days of June 2018.\par

The analysis is carried out by developing computer codes to handle high-resolution voluminous data for sorting, compressing, averaging and generating result oriented plots for visualisation, intercomparison and depicting scientific phenomena. \par

\section*{Results and Discussion}

\begin{figure}[h]
\centering
{\includegraphics[width=\textwidth]{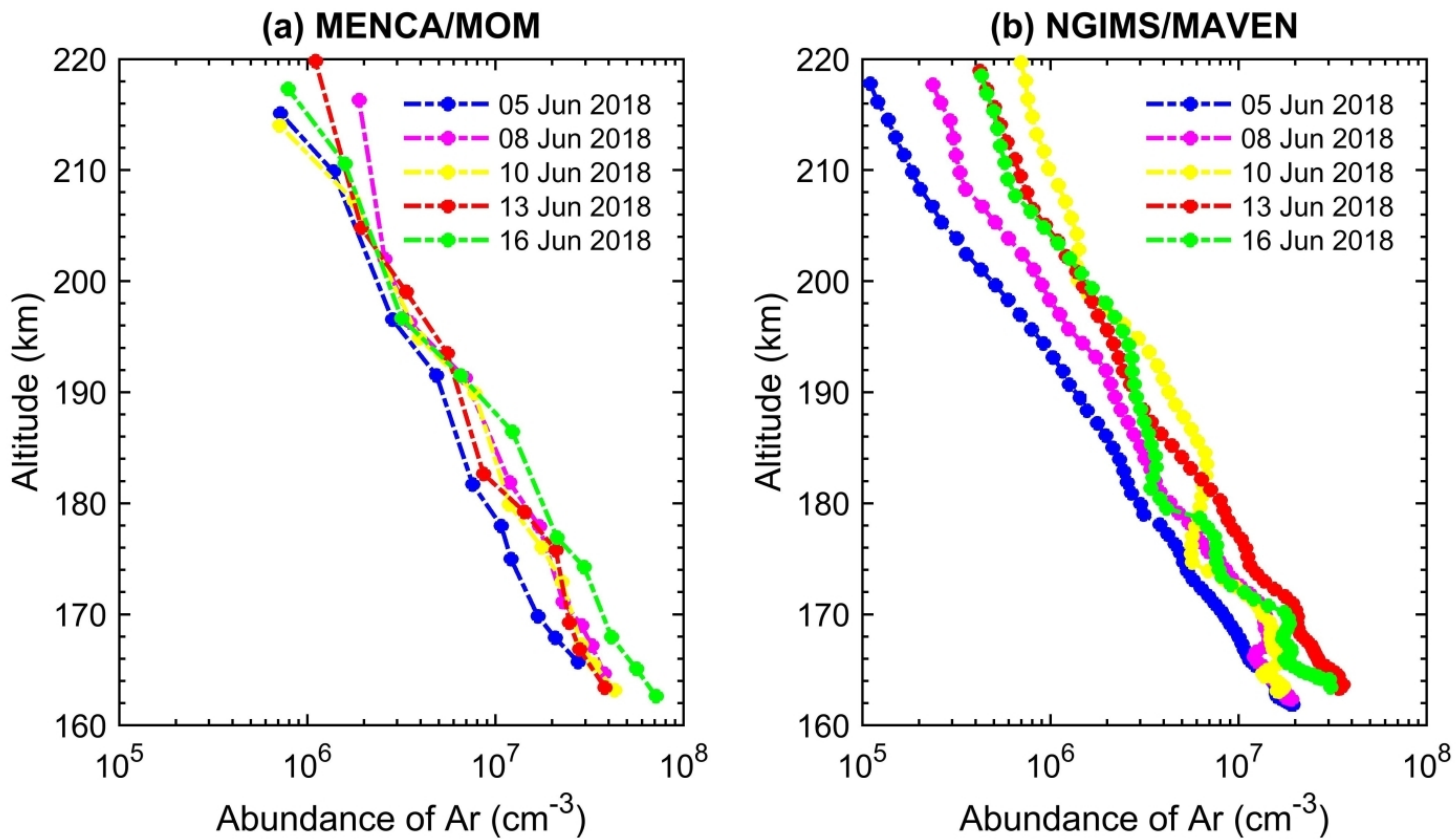}}
\caption{Mean denity profiles of Argon in the thermosphere of Mars as measured by NGIMS/MAVEN and MENCA/MOM during a few selected days of June 2018.}
\label{fig001}
\end{figure}

The repetitive coverage of altitudes near MOM's periareion is less frequent than that of MAVEN due to large differences in orbital periods of about $\approx$67\,hr and $\approx$4.5\,hr respectively. The mean density profiles of Argon ($Ar$) measured by both the spacecraft on a few typical days during June 2018 are shown in figure \ref{fig001} for comparison. The MENCA data on the Argon profile were retrieved from the published results of \citet{Rao2020EnhanceDensity_MAVEN_MOM} on the effect of PEDE on Mars thermosphere, as mentioned earlier.

The figure shows the thermospheric variation of $Ar$ density up to the exobase. While the primary trend of the density profiles for the four selected days measured by NGIMS and MENCA are comparable, the absolute values are understandably different due to the measurements carried out at two opposite points of the dawn-dusk terminator. The slopes of the profiles are similar to the exobase as can also be derived using equations \ref{eq001} and \ref{eq002} with modelled thermospheric temperature structure. Significant day-to-day variation of the overall thermospheric density is observed, and it is found that $Ar$ densities are lower in the early days of June 2018 than towards the middle. Except for its ionisation by solar EUV radiation ($ Ar + h\nu \rightarrow Ar{^+} + e $) ; $Ar$ is considered to be a stable constituent and may only vary significantly due to any temperature changes brought about by variations in solar EUV radiation, gravity wave dissipation and dust storm-driven aerosol warming. \par

\begin{figure}[h]
\centering
{\includegraphics[width=\textwidth,height=6.5cm]{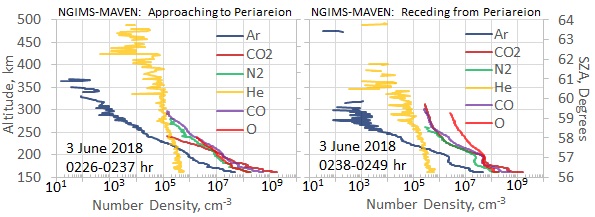}}
\caption{Density profiles of gas constituents in the thermosphere/exosphere of Mars as measured by NGIMS/MAVEN during morning hours of the local solar time of 03 June 2018 for either side of passage through the periareion altitude.}
\label{fig002}
\end{figure}

Since the heating effect of the PEDE peaked towards the end of June 2018, we further explore the state of the natural and photochemically produced constituents from the thermosphere into the exosphere during June 01-15, 2018. Here we search for effects other than solar activity, particularly the solar wind plasma energetics, which is variable on a day-to-day basis. The altitude profiles of densities of $CO_{2}$, $Ar$, $N_{2}$, $CO$, $O$, and $He$ are shown for single MAVEN orbit on 03 June 2018, in figure \ref{fig002}.

It can be seen from the figure \ref{fig002} that there are differences in profile shapes for the two epochs with post-periareion curves (frame on the right-side), indicating more extended scale heights in the exospheric region due to the increase in the mean free path of the gases. In further analysis, we use the post-periareion part of the measurement.\par

\begin{figure}[h]
\centering
{\includegraphics[width=\textwidth,height=10cm]{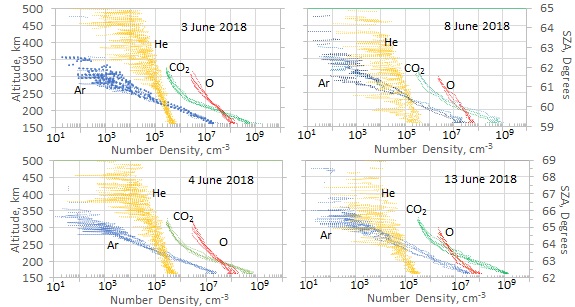}}
\caption{Density profiles of some gas constituents on a few selected days of June 2018 for all available orbital traverses of MAVEN/NGIMS through the periareion altitude.}
\label{fig003}
\end{figure}

During June 01-15, 2018, on average, 4-5 passes of MAVEN per day have produced useful data with minimal variation of solar zenith angle, local solar time, latitude and longitude from one pass to another and hence it is well suited to study short term variations of density profiles. Figure \ref{fig003} shows the result of such an analysis for four selected days to highlight the diurnal as well as the day-to-day variation of the density profiles of $Ar$, $He$, $CO_{2}$ and $O$.\par

The figure \ref{fig003} shows that the gas concentrations have significant diurnal variation, but the day-to-day changes are also present, evidenced by a relatively larger diurnal spread of density variations on 08 and 13 June 2018. The elevation of the $CO_{2}-O$ cross-over altitude from an average of 200\,km to $\approx$230\,km on 13 June 2018 is quite prominent in the figure. We need further to study this day-to-day effect and its possible causes. Since the early 1970's 1D photochemical and 3D general circulation based models of the thermosphere-exosphere of Mars are being developed and continuously improved using new observations from space probes, including landers and orbiters, these models provide the values of neutral and ionic composition, densities and temperatures with possible modulation for diurnal, seasonal and solar cycle variation \citep{Krasnopolsky2002MarsAtmos_LowMedHighSolarActivity,Bougher2015MarsIonoThermoModel_SolarDiurnalSeasonalCycleMUA}. While these models are potentially suited for projecting mean day-to-day variation brought about by solar wind plasma, EUV, etc., these would need verification with observed data.  Here we emphasise to explore this aspect of the Martian phenomenon from the relevant \emph{in-situ} measurements. \par  

\begin{figure}[h]
\centering
{\includegraphics[width=\textwidth,height=8cm]{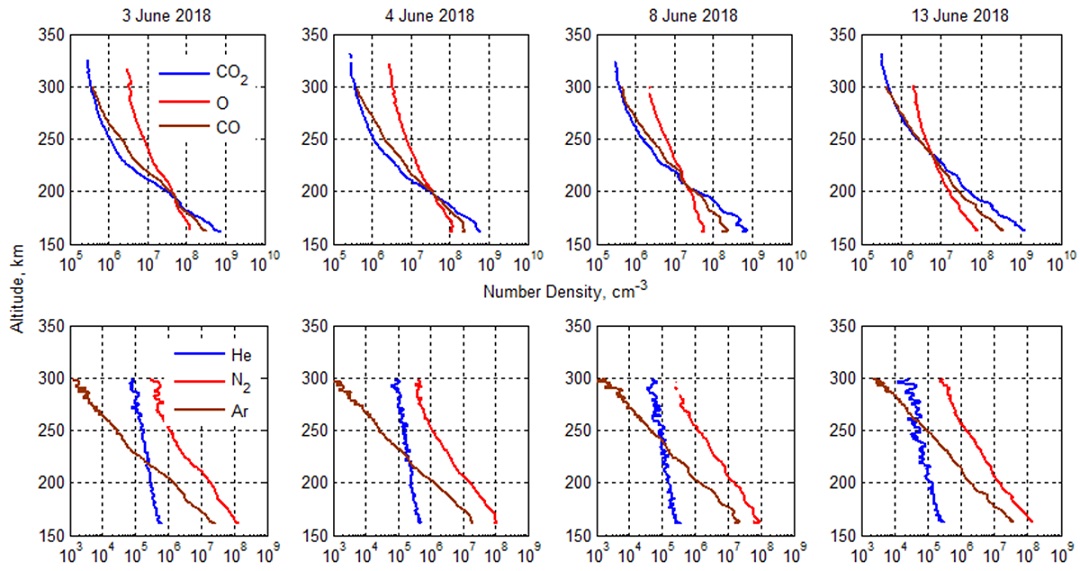}}
\caption{Daytime mean density profiles of a few parent and photochemically produced atmospheric constituents for four seleted days during June 2018 using MAVEN/NGIMS data covering Mars' thermosphere and lower exosphere.}
\label{fig004}
\end{figure}

Figure \ref{fig004} shows average daytime profiles of a few crucial parents and photochemically produced constituents in the thermosphere-exosphere of Mars during the selected days of June 2018.\par

\begin{figure}[h]
\centering
{\includegraphics[width=\textwidth,height=8cm]{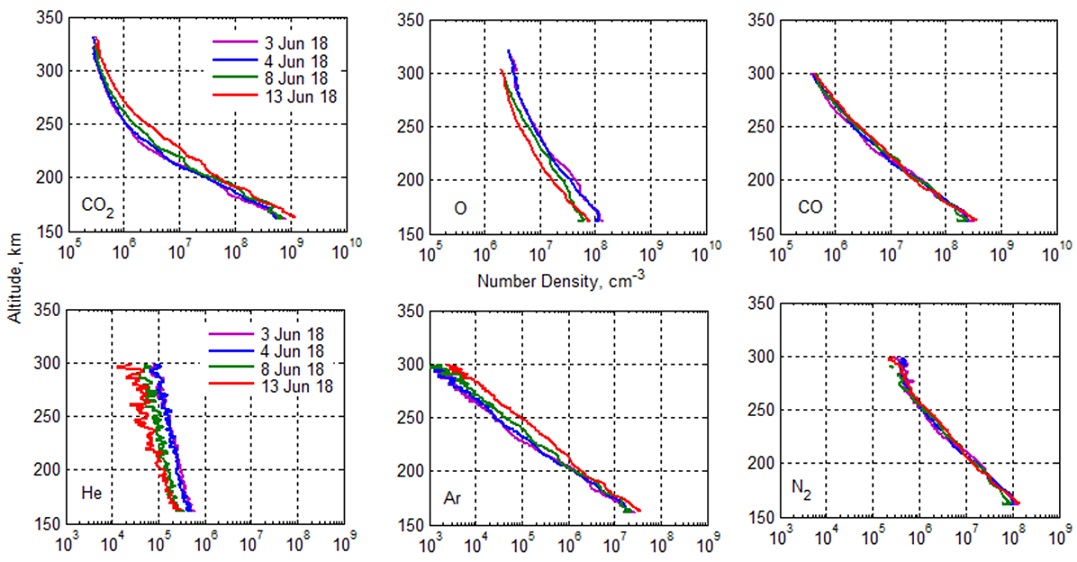}}
\caption{Daytime mean density profiles of a few thermospheric/exospheric constituents during the same four selected days of June 2018 using MAVEN/NGIMS observations.}
\label{fig005}
\end{figure}

The constituents selected in the plots are in two groups, the first with $CO_{2}$, $O$ and $CO$; the other with $He$, $N_{2}$ and $Ar$. It is interesting to see that over the short interval between 03-13 June 2018, except for $N_{2}$ and $CO$, all other constituents show a shift of profile density values either towards an increasing or decreasing trend. For example, $CO_{2}$ increases and $O$ decreases; $He$ decreases, but $Ar$ increases as we go from 03 to 13 June 2018 in time. These results indicate the competing effects of change in thermospheric temperature and efficiency of photochemical reactions discussed earlier. More details of the variations of individual constituents are given in figure \ref{fig005}.

\begin{figure}[h]
\centering
{\includegraphics[width=\textwidth,height=12cm]{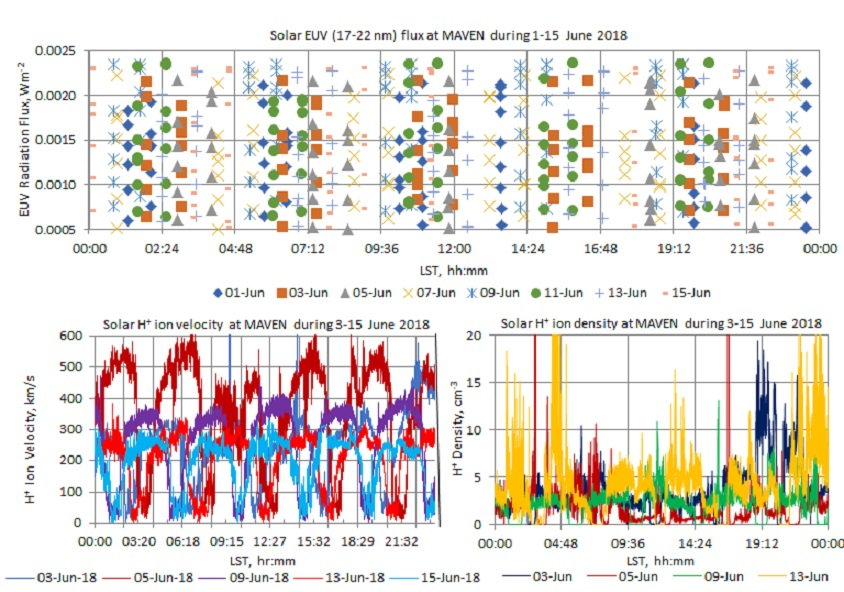}}
\caption{\textbf{Top Panel}: Solar EUV radiation fluxes for the channel 17-22\,nm with 1\,nm resolution and two values during each orbit to cover the 24\,hr period of each day during selected days of June 01-15, 2018; \textbf{Bottom (Left Panel)}: Solar wind proton velocities at MAVEN orbit to cover the 24-hour duration of the same days; \textbf{Bottom (Right Panel)}: Solar wind proton fluxes at MAVEN for the same hours and days.}
\label{fig006}
\end{figure}

\begin{figure}[h]
\centering
{\includegraphics[width=\textwidth]{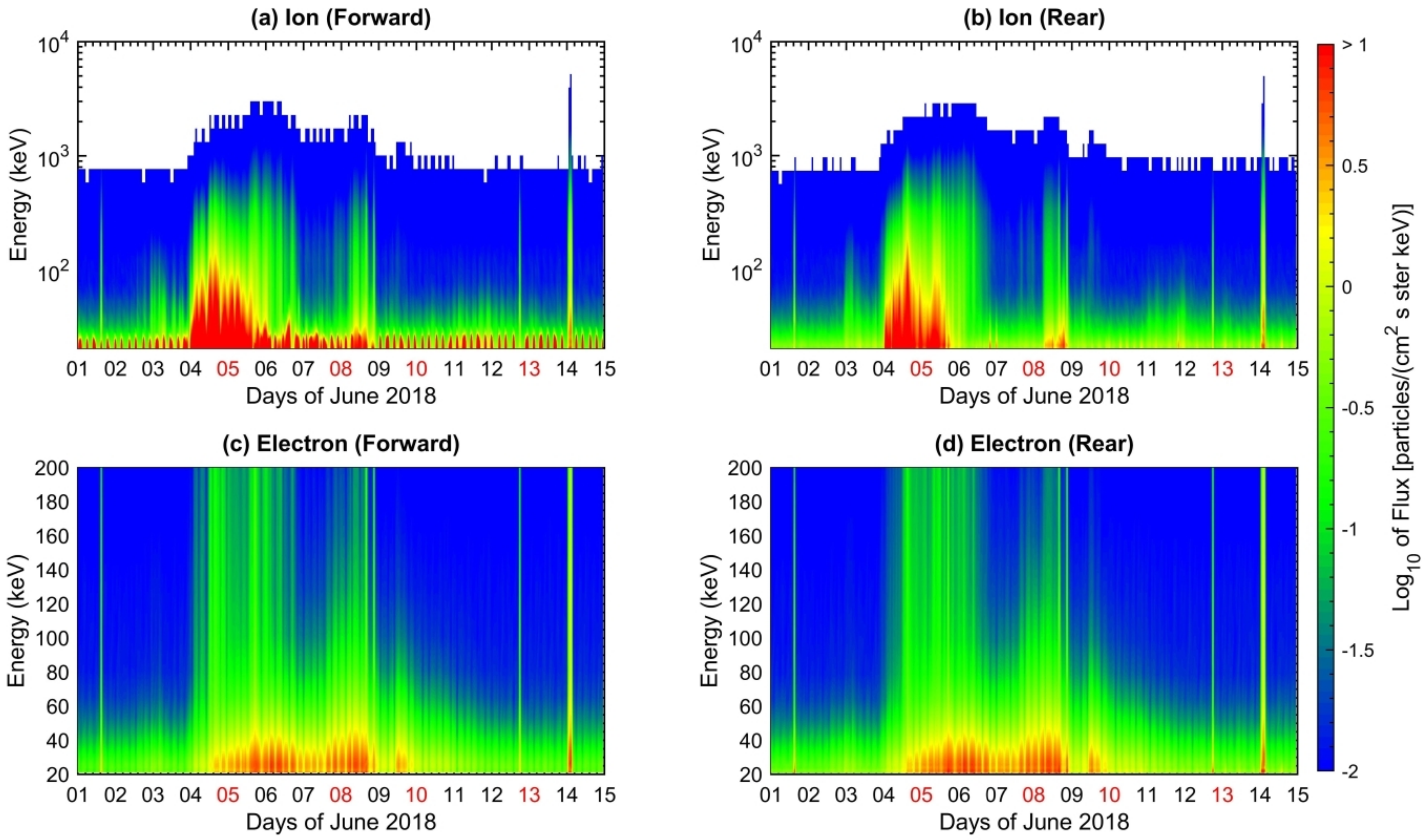}}
\caption{Solar proton/Hydrogen-ion velocities and integrated fluxes for canonical Parker spiral directions signifying solar wind variation at MAVEN orbit during 04-14 June 2018.}
\label{fig007}
\end{figure}

It can be noted that there is only a minimal variation of the density profiles of $N_{2}$ and $CO$ as compared to $CO_{2}$, $O$, $He$ and $Ar$. The direction of change in densities is also reversed between $CO_{2}$-$O$ and $He$-$Ar$. Many inferences can be drawn from this first of its kind result. Before going into the details of interpretation, we plot the solar EUV radiation fluxes, solar wind proton velocities and fluxes using the EUVM and SEP instruments' data onboard MAVEN during the same period the first half of June 2018 as shown in figures \ref{fig006} and \ref{fig007}.

The combined results of figures \ref{fig006} and \ref{fig007} show that while, on average, the solar EUV radiation has not changed, the solar electron/proton velocities show a steady decrease between 01-15 June 2018. It can be seen from figure \ref{fig007} that the solar EUV radiation fluxes for each observation scan are a function of wavelengths between 17 to 22\,nm, so there are six values in one wavelength scan, which are renewed in time after two such scans with the change in the orbit of MAVEN covering the whole day. These sets of six EUV values corresponding to each wavelength in a set are not seen to be changing between different days of the period 01-15 June 2018. The two lower panels of figure \ref{fig006} and the four sub-plots of figure \ref{fig007} show a definite decreasing trend of the solar wind proton velocities and fluxes from beginning to middle of June 2018. This aspect of considerable variations in solar wind particle kinetic energy without similar variations of the solar EUV radiation provides exciting insights to the results on density changes shown in figures \ref{fig004} and \ref{fig005}.\par

In the 1D atmospheric model developed by \citet{Krasnopolsky2002MarsAtmos_LowMedHighSolarActivity}, many possible photochemical and ion-chemical reactions have been considered. The model provides variations of thermospheric constituents, some of which have been measured and compared under similar solar and geometric conditions. In the background of this model, for our purpose to explain the special event, we consider the parent species as $CO_{2}$, $N_{2}$, $Ar$, $He$ and their photochemical products as $O$ and $CO$. Since here we are not directly dealing with the escape of gases from Mars, we have not considered Hydrogen, Deuterium and related species. Similarly, the primary ions taken here are $CO_{2}^{+}$, $O^{+}$, $CO^{+}$, $N_{2}^{+}$, $Ar^{+}$ and $He^{+}$. In the first row of figure \ref{fig005}, the variations of $CO_{2}$, $CO$ and $O$ indicate that while $CO$ concentration in the upper thermosphere extending to the exosphere has remained nearly constant during the period of study, $CO_{2}$ and $O$ show clear increasing and decreasing trends in the density values respectively.\par

Since these variations have taken place when solar EUV radiation has remained nearly constant, and the PEDE effect started significantly towards end of June 2018, the differences are attributable to the distinctly changing characteristics of the solar wind charged particle radiation energy fluxes. As explained by \citet{Bougher2015MarsIonoThermoModel_SolarDiurnalSeasonalCycleMUA}, the solar wind plasma can provide electron impact energy to dissociate or ionise the gas constituents in the Martian thermosphere/exosphere. Taking this energy input from electrons/protons, the following sequence of reactions (extending the set of equations \ref{eq003} to \ref{eq007} above) would result in enhanced $O$ and reduced $CO_{2}$ concentrations as given below:

\begin{eqnarray}
O{^+} + CO{_2} \rightarrow O{_2^+} + CO \label{eq011} \\
CO{_2^+} + O \rightarrow O{_2^+} + CO \label{eq012} \\
O{_2^+} + e \rightarrow 2O \label{eq013} \\
CO + e{^*} \rightarrow CO{^+} + e + e{^*} \label{eq014} \\
CO + e{^*} \rightarrow C{^+} + O + e + e{^*} \label{eq015} 
\end{eqnarray}

These interactions initiated by solar wind energetic plasma lead to different densities of $O$ and $CO$, as shown in equations \ref{eq011}--\ref{eq013}. However, $CO$ further dissociates (equations \ref{eq014} and \ref{eq015}), and hence there is no net change in its concentration as expected with decreasing values of solar wind plasma energy and fluxes. Hence the steady increase of $CO_{2}$, a decrease of $O$ and relative constancy of $CO$ densities with the decrease of solar wind plasma velocities/fluxes between the beginning till the middle of June 2018 observed by NGIMS/MAVEN can be explained by the set of 15-equations given in this paper. This short period solar wind plasma variation and the combined response of $CO_{2}$, $O$ and $CO$ is a first-time result. This indicates that in addition to the density changes between minimum to maximum solar cycle predicted by various models, the day-to-day solar wind variations of plasma energy and flux also produce observable density variations of Martian thermospheric and exospheric gaseous concentrations, and these should be detectable in future models predictions. Unlike on Earth, this phenomenon is possible due to the absence of a global magnetic field of Mars, enabling direct interaction of solar wind charged particles with upper atmospheric gases. \par

Coming back to the four plots of the first row of figure \ref{fig004}, we notice that the cross-over altitude between $CO_{2}$ \& $O$ density profiles progressively moves up from $\approx$190 to $\approx$240\,km during 03-13 June 2018. This can be realised from the effect explained for figure \ref{fig005} where $CO_{2}$ concentration increased and that of $O$ decreased for the same period resulting in the cross-over altitude's upward movement. \par

The plots in the second row of figure \ref{fig005} show that $N_{2}$ densities are not appreciably affected by the solar wind plasma variations for the days under study. We note the following possible ionisation reactions involving $N_{2}$ :

\begin{eqnarray}
N{_2} + e{^*} \rightarrow N{_2^+} + e{^*} + e \label{eq016} \\
N{_2^+} + CO{_2} \rightarrow N{_2} + CO{_2^+} \label{eq017} 
\end{eqnarray}

The regeneration of $N_{2}$ through equation \ref{eq017} is very fast, and hence even if $N_{2}$ gets ionised by energetic electrons, its overall density remains almost constant throughout the period. \par

To understand the observed decreasing $He$ and increasing $Ar$ profile trends shown in figure \ref{fig005}, we consider the following reactions:

\begin{eqnarray}
He + e{^*} \rightarrow He{^+}{^*} + e{^*} + e \label{eq018}\\
Ar + e{^*} \rightarrow + Ar{^+} + e{^*} +e \label{eq019}\\
Ar{^+} + CO{_2} \rightarrow  Ar + CO{_2}{^+} \label{eq020}
\end{eqnarray}

where $\boldsymbol{He{^+}{^*}}$ denotes energetic $\boldsymbol{He}$ ion.

Helium is a very light gas constituent, gets not only ionised by solar wind plasma but also gets energised to enhance its kinetic energy, resulting in higher densities as inferred from equations \ref{eq001} and \ref{eq002}. The temperature rise would lead to higher pressure and density, assuming that the velocities are still less than Helium's escape velocity. On the other hand, Argon gets ionised to produce $Ar$ ions, which amount to a loss to $Ar$ concentration even though only some of them get converted into $Ar$ atoms with a relatively slow reaction rate (equation \ref{eq020}). So like $CO_{2}$, $Ar$ shows an increase in density in the tapering of solar wind particle radiation energetics. Now getting into figure \ref{fig004} again, the rise of the cross-over altitude between $He$ and $Ar$ densities is caused because $Ar$ density increases, but $He$ density decreases as the solar wind energy and flux decrease during 03-13, June 2018.\par

So it is possible that the day-to-day changes in solar wind plasma can affect the thermosphere/exosphere density of gaseous constituents in a significant way.

\section*{Conclusion}

Measurements made using the mass spectrometric technique onboard MAVEN and MOM orbiter missions around Mars for $Ar$ gas in the thermosphere/exosphere indicate variations of its density profiles within a short period of 15-days during June 01-15, 2018. This study has been extended by using further data from NGIMS/MAVEN on the variations of parent species ($CO_{2}$, $N_{2}$, $Ar$ and $He$) as well as photochemical products $O$ and $CO$. The solar EUV flux and solar wind plasma velocity and flux data were obtained for the same period from the scientific experiments EUVM and SEP onboard the MAVEN probe orbiting Mars. Because the solar wind plasma velocities and fluxes showed considerable enhancement at the beginning of June 2018 and later returned to typical average values by the middle of June 2018. In comparison, the solar EUV fluxes ($\lambda$= 17-22\,nm) remained nearly constant during the same period; possible effect due to this solar energetic plasma variation alone on the thermosphere/exosphere density profiles measured by NGIMS payload was examined. The results show that the energetic plasma of solar wind affected $CO_{2}$, $O$, $Ar$ and $He$ but $N_{2}$ and $CO$ densities remained nearly constant. While $CO_{2}$ and $Ar$ densities showed increasing trends in the thermosphere/exosphere region during June 03-13 (concurrent with the decreasing kinetic energy and density of solar wind plasma), the reverse was the case for $O$ and $He$. In other words, with the enhanced solar wind's energy and density during the first week of June 2018, $CO_{2}$ and $Ar$ densities were lower compared to their densities in the second week of June 2018. The reverse is the case for $O$ and $He$. We have tried to interpret this new result by considering the role of a large number of electron impact dissociation and ionisation reactions initiated by the solar wind plasma. The observed variation of the cross-over altitude between $CO_{2}$ and $O$ profiles and $Ar$ and $He$ profiles are also explained based on these theoretical considerations. Future models providing the structure and composition of Martian thermosphere/exosphere would need to incorporate computations of solar wind plasma driven composition changes in addition to the solar EUV related variations.

\section*{Acknowledgements}

This work was funded by the Indian Space Research Organisation (ISRO) under Mars Orbiter Mission's Announcement of Opportunity Program through the research project, Observation and Modeling Studies of the Atmospheric Composition of Mars (OMAC), vide ref: ISRO:SPL:01.01.33/16. We greatly acknowledge the use of MENCA data from MOM, archived at Payload Operation Center, Space Physics Laboratory, Vikram Sarabhai Space Centre, Thiruvananthapuram, India and at the Indian Space Science Data Center (ISSDC), Bengaluru, India \url{https://www.issdc.gov.in/}. The NGIMS, SWIA, EUVM and SEP datasets of MAVEN used for this study were publicly available on MAVEN Science Data Center at LASP \url{https://lasp.colorado.edu/maven/sdc/public/} as well as the Planetary Data System \url{http://pds.nasa.gov}. The MAVEN mission is supported by NASA through the Mars Exploration Program. This research work was carried at Atmospheric and Space Science Research Lab, Department of Physics, Bangalore University, Bengaluru, India.

\bibliography{ref.bib}
\bibliographystyle{apalike}

\end{document}